\documentclass[journal]{IEEEtai}

\usepackage[colorlinks,urlcolor=blue,linkcolor=blue,citecolor=blue]{hyperref}

\usepackage{color,array}
\usepackage{xcolor}

\usepackage{graphicx}

\usepackage{orcidlink}

\begin{document}

\title{Cyber Shadows: Neutralizing Security Threats with AI and Targeted Policy Measures} 

\author{Marc Schmitt \orcidlink{0000-0003-4550-2963}, Pantelis Koutroumpis \orcidlink{0000-0002-2398-0027} 
\thanks{Forthcoming in IEEE Transactions on Artificial Intelligence.}
\thanks{M. Schmitt is with Siemens AG, Munich, Germany. The views expressed in this article are solely those of the author and do not necessarily reflect the views of Siemens AG.
}
\thanks{P. Koutroumpis is with the Oxford Martin School, University of Oxford, Oxford, U.K.}}

\maketitle

\begin{abstract}
The digital age, driven by the AI revolution, brings significant opportunities but also conceals security threats, which we refer to as cyber shadows. These threats pose risks at individual, organizational, and societal levels. This paper examines the systemic impact of these cyber threats and proposes a comprehensive cybersecurity strategy that integrates AI-driven solutions, such as Intrusion Detection Systems (IDS), with targeted policy interventions. By combining technological and regulatory measures, we create a multilevel defense capable of addressing both direct threats and indirect negative externalities. We emphasize that the synergy between AI-driven solutions and policy interventions is essential for neutralizing cyber threats and mitigating their negative impact on the digital economy. Finally, we underscore the need for continuous adaptation of these strategies, especially in response to the rapid advancement of autonomous AI-driven attacks, to ensure the creation of secure and resilient digital ecosystems.
\end{abstract}

\begin{IEEEImpStatement}
This paper contributes to the ongoing efforts to build a safer digital world by addressing the growing cybersecurity challenges posed by artificial intelligence. We propose a strategy that combines AI-driven security technologies with policy measures to protect individuals, businesses, and society from evolving cyber threats. Our work aims to not only neutralize these threats but also safeguard essential values like privacy, fairness, and security in the digital economy. By offering practical tools and recommendations, including a 'potential threat directory,' we help policymakers, researchers, and organizations better prepare for future risks and coordinate responses. The ultimate goal is to foster secure and resilient digital ecosystems that can adapt to the rapidly changing landscape of AI-driven cyberattacks.
\end{IEEEImpStatement}


\section{Introduction}

\IEEEPARstart{A}{rtificial} Intelligence (AI) has reached human level performance across a wide range of knowledge domains, due to a mix of computing power, data availability and algorithmic innovations \cite{lecun2015deep}. A combination of these core capabilities has allowed applications to foster the invention of new algorithms \cite{mankowitz2023faster}, advance medical treatments \cite{moor2023foundation}, augment human creativity \cite{jia2023and}, and accelerate scientific progress \cite{sourati2023accelerating}. Scaling laws predict further improvements in the near term that could lead to general-purpose AI agents for a range of technological applications.

While the upsides of these new capabilities can boost productivity in several segments and markets, their potential downsides have drawn the attention of scholars and policymakers.
At the core of these downsides lies the "trust" that users can have when interacting with LLMs. In this process several recent works have pointed to serious flaws. First, Large language models can be easily distracted by irrelevant context  which significantly reduces the consistency and performance \cite{distraction_ICML23}. Second, artificial agents have traditionally been trained to maximize reward, which may incentivize power-seeking and deception, analogous to how next-token prediction in language models (LMs) may incentivize toxicity. This often leads to harmful and anti-social behavior that needs to be mitigated during training  \cite{machiavelli_ICML23}. Beside these, researchers have also uncovered that the traditional Turing test may not be enough to validate the performance of LLMs as an arbitrary person, which led them to point to the hyper-accuracy distortion, uncovering a trait that is not common when real humans respond to specific questions \cite{replication_ICML23}.

\begin{figure*}[ht]
\vskip 0.2in
\begin{center}
{\includegraphics[width=1\textwidth]{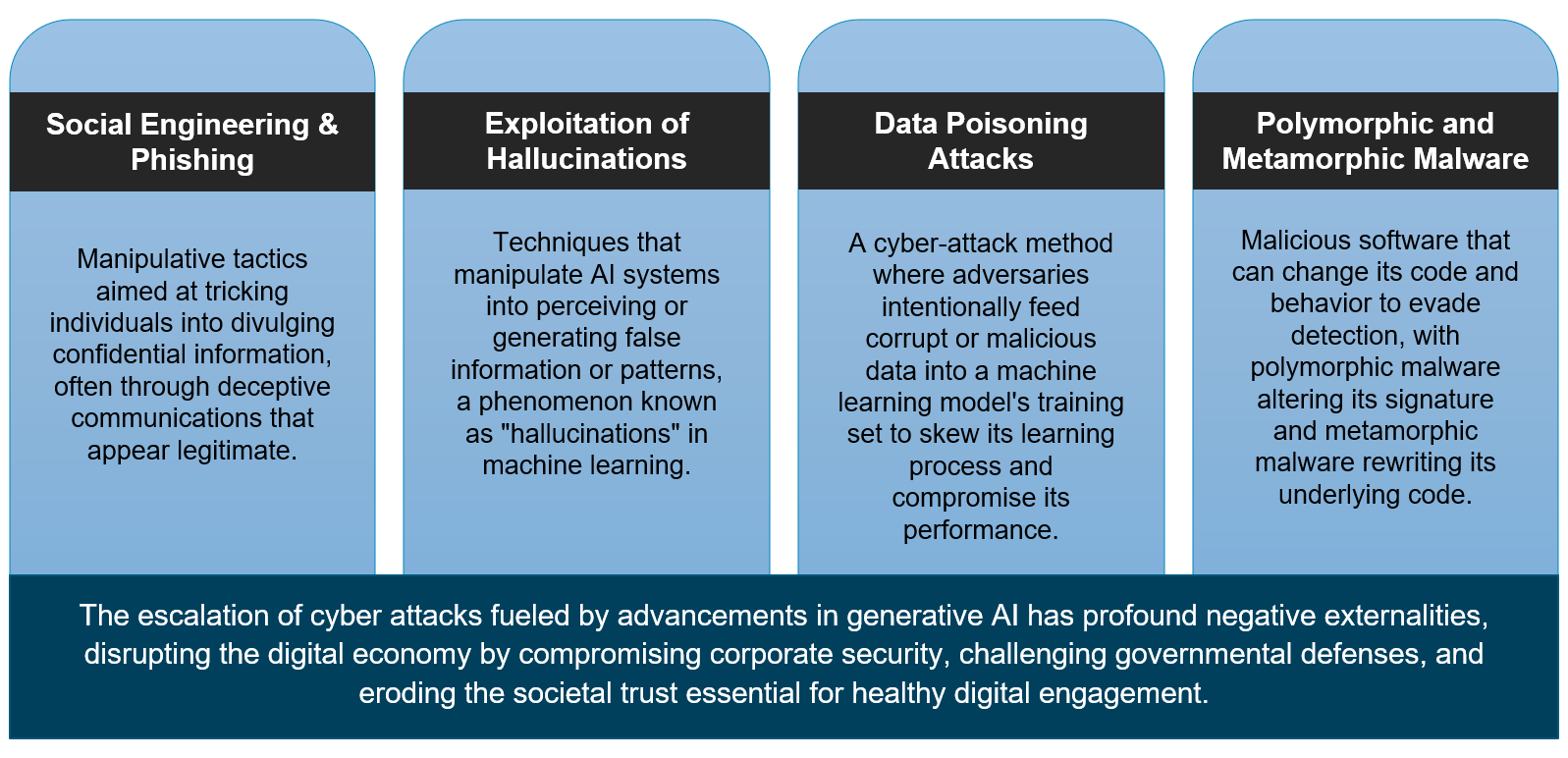}}
\caption{Technology driven Threat Amplification by Generative AI}
\label{icml-historical}
\end{center}
\vskip -0.2in
\end{figure*}

Policymakers have also responded to these first signs. The UK Safety Summit initiated a global exchange of policies around the governance of AI and coined the term “frontier AI” to capture the importance of this undertaking\footnote{“Highly capable general-purpose AI models that can perform a wide variety of tasks and match or exceed the capabilities present in today’s most advanced models.”}. A few days prior to the UK Summit the US President signed an executive order on AI, setting high standards for AI safety and security, safeguarding privacy, ensuring equity and civil rights, and promoting innovation\footnote{ \href{https://www.federalregister.gov/documents/2023/11/01/2023-24283/safe-secure-and-trustworthy-development-and-use-of-artificial-intelligence}{US AI Executive Order}}. The European Union was the first to start the work on its EU AI Act\footnote{ \href{https://artificialintelligenceact.eu/the-act/}{EU AI Act}}, pointing towards a risk-based regulatory framework.

In this paper we focus on the ways that the increasing use of AI can amplify the exposure to cyber-threats either directly to its users or through negative externalities to individuals and organizations not involved in those activities. Additionally, we will delve into strategies for neutralizing these cyber shadows through the integration of AI-driven security solutions and targeted policy measures.

\section{Cyber Shadows}
\label{Shadows}

The proliferation of Generative AI escalates existing threats and brings novel vulnerabilities. Cyber shadows refer to the hidden or amplified security threats that emerge within digital ecosystems due to the use of advanced AI technologies. These threats may be direct, as seen in AI-driven attacks that enhance traditional cyber threats, or indirect, that impact the wider digital ecosystem. For these latter indirect threats, we borrow the term "negative externalities" from economics, to refer to unintended and often widespread consequences that affect third parties or the system at large, such as data breaches or the erosion of trust in digital interactions. Due to their human-level text generation capabilities, GenAI can be used to launch social engineering attacks in all types of interactions, including computer-to-computer, human-to-computer, and human-to-human. Thus, LLMs pose a notable threat to the cybersecurity landscape. These AI systems can be exploited to create sophisticated social engineering and phishing attacks, produce realistic forgeries, or automate the generation of malicious code at a scale and sophistication previously unattainable \cite{Schmitt2024}. LLMs can be accessed with techniques like jailbreaking, prompt injection, and reverse psychology, which are attack attempts to manipulate a system by altering its usual behavior. AI-enabled agents have the potential to automate various parts, if not the entire lifecycle (cyber kill chain), of cyber attacks. In addition, AI agents can – theoretically – learn from each attack and improve their success rate. In this context, individuals are more likely to forfeit control of their own accounts and as a result jeopardize their own and other’s data. As AI/ML technologies have the potential to both detect and amplify wide range of cyber attacks it is essential to understand the nature of such techniques and develop proactive strategies to mitigate this escalating threat.

\subsection{Native AI Threat Amplification}
\label{Threats}

In this section we outline the technology driven threats which represent the direct cyber shadows for AI systems. This means that we are focusing on the "off the shelf" characteristics of LLMs and not on their misuse. In Section \ref{Externality} we look into the threats from LLM misuse and also present further indirect effects (externalities)  that derive from its use.

\subsubsection{Automated Code Creation}
\label{Code}

\begin{figure}[ht]
\begin{center}

{\includegraphics[scale=0.58]{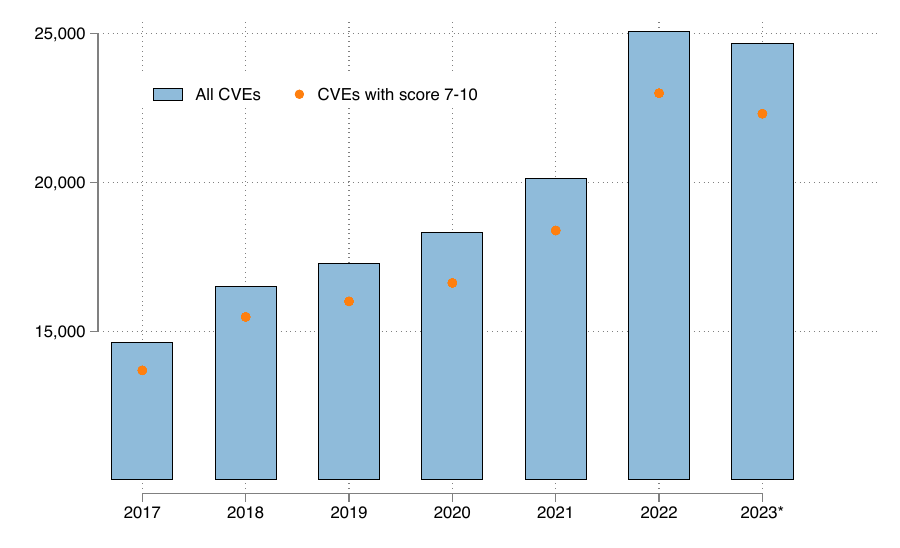}}
\caption{Vulnerability and Exposure Increase, Source: National Vulnerability Database (NVD/NIST), data for 2023 as of 11th November 2023. CVSS scores range from 0-10 with 10 representing the most critical CVEs.}
\label{icml-historical2}
\end{center}
\end{figure}

The integration of AI into critical infrastructure without comprehensive security measures could lead to new attack vectors, potentially destabilizing essential services. During the last decade, there has been a consistent increase in the number of common IT security vulnerabilities and exposures (CVEs) worldwide and a more pronounced rise in the past two years both in absolute numbers and severity (see Figure \ref{icml-historical2}).

Generative AI through its ability to automate code generation/creation has led to a phenomenal rise in the usage of AI assistants with GitHub’s Co-pilot “behind an average of 46 Percent of the cross-language code written. As the number of CVEs continues to grow fast, users of assisted code have been found to “believe they wrote secure code than those without access to the AI assistant” when in reality those “who trusted the AI less and engaged more with the language and format of their prompts […] provided code with fewer security vulnerabilities” \cite{perry2023users}. Further research has shown that both ChatGPT “often generates source code that is not robust to well-known attacks” \cite{khoury2023secure} and GitHub’s Co-pilot is “more likely to generate vulnerable code in response to prompts that correspond to older vulnerabilities.” \cite{asare2023github} 

Acknowledging the cyber threat risks in code generation in a recent (27th June) workshop organized by Google, Stanford University, and the University of Wisconsin-Madison participants agreed in a joint document that there is a lack of comprehensive code-related LLM capabilities which is necessary “to inform possible defenses and possible threats”. There is ongoing research on security hardening techniques that guide code generation towards these constantly changing security needs “without modifying the LM's weights” which shows promise but requires further effort \cite{he2023large}. In particular these security hardening techniques largely rely on fine-tuning existing models to make them adapt to new code vulnerabilities. However, there is a separate push within this area that tries to prevent malevolent actors from fine-tuning existing modes \cite{deng2024sophon}. This approach makes fine-tuning as costly - in terms of compute - as training from scratch and prohibits a quick response in cases where new vulnerabilities emerge. The "balance of incentives" between safety and cybersecurity renders a quick solution less obvious and should alert policy-makers in terms of the limits of automated coding generation.

\subsubsection{Social Engineering and Phishing}
\begin{figure*}[ht]
\vskip 0.2in
\begin{center}
{\includegraphics[width=1\textwidth]{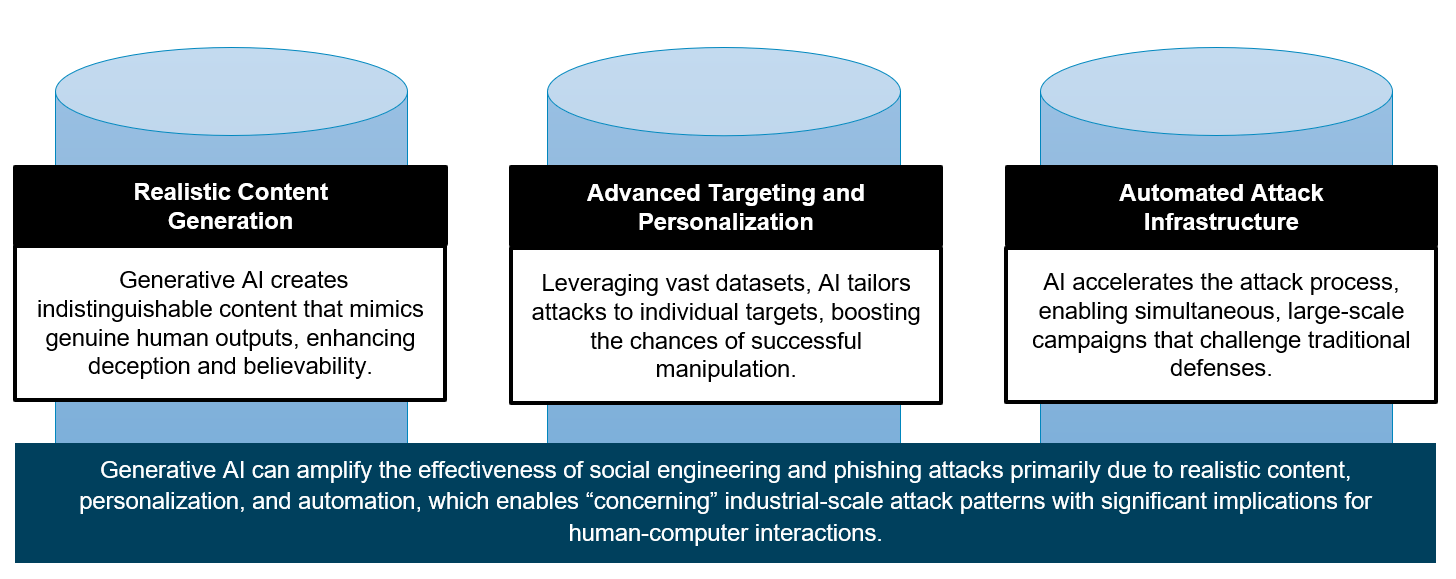}}
\caption{The Three Pillars of GenAI induced Threat Amplification in Social Engineering \cite{Schmitt2024}}
\label{SE}
\end{center}
\vskip -0.2in
\end{figure*}

The impact of generative AI on SE can be split into three major pillars:

\textbf{Content Generation:} Generative AI's foremost capacity lies in creating convincing digital content, critical in phishing scams like website cloning, where AI can duplicate and subtly alter legitimate websites to trick users (see Figure \ref{SE}). It excels in generating persuasive texts, authentic-looking images, credible voice recordings, and deepfake videos that mimic real individuals, elevating the threat of misinformation and deception. This technology can be misused for impersonation, bypassing biometric security, extortion through fabricated content, or spreading disinformation. 

\textbf{Personalization:} Generative AI accelerates the shift toward hyper-personalized cyber attacks by enabling attackers to craft highly convincing content and strategies tailored to individuals' online behaviors and profiles. This AI-facilitated reconnaissance gathers detailed intelligence to inform pretexting, where attackers create credible narratives for engagement. Consequently, the execution of attacks becomes more sophisticated, leveraging AI-generated emails, messages, or interactions that convincingly mimic known contacts or organizations, thus increasing the likelihood of successful manipulation.

\textbf{Scalability:} AI-driven automation empowers attackers to execute widespread and sophisticated phishing and social engineering campaigns with little effort. By analyzing victims' behaviors, AI enhances the convincing nature of attacks, allowing for real-time adaptation in response strategies through feedback loops. Intelligent bots within AI-powered botnets can autonomously engage with potential victims on a large scale, simulating human interactions and adapting messages based on feedback. Moreover, generative AI can craft content that bypasses security measures, using polymorphic techniques to evade detection (Section \ref{poly}). This level of automation allows for unprecedented scale and efficacy in cyber attacks, challenging current cybersecurity defenses.

\subsubsection{Exploitation of Hallucinations}

The exploitation of LLMs' hallucinations has recently been shown to be the target of potential attackers. As code-assistants and chat-bots often generate responses for URLs that do not exist in reality, attackers can leverage these responses and host their malicious code in these links. Once a user receives these responses from the AI-assistant, they will download the malicious code. The probability of receiving a hallucinated response was more than 20\% for Node.js (40 cases with at least one package that has not been published out of 201) and more than 35\% for Python packages (80 responses with at least one unpublished package out of 227 questions)\footnote{The source and details of these attacks can be found in this link: https://vulcan.io/blog/ai-hallucinations-package-risk}.

\subsubsection{Data Poisoning Attacks}

Recent progress in the field has led to the increasing incorporation of machine learning models into a variety of practical applications, aiding humans in their everyday decision-making processes \cite{Schmitt2023DeepReality, Eapen2023HowCreativit, Shepherd2022MachinesEntrepreneurship}. Consequently, the rise of these applications makes data poisoning attacks a significant problem. These attacks, wherein maliciously altered data are fed into the training set of a model, can severely compromise the integrity and reliability of machine learning systems, potentially leading to erroneous or biased decisions in critical applications. Multimodal models, which use both visual and linguistic data, have recently become prominent but face the risk of poisoning attacks where adversaries manipulate training data to induce harmful behaviors. \cite{poisoning_ICML23} examined such attacks in both visual and linguistic domains, aiming to determine their relative vulnerability. The authors developed three types of poisoning attacks for multimodal models and tested them across various datasets and architectures. The results show that these attacks are effectively harmful in both modalities, although the extent of their impact varies. In response, the authors devised defense strategies for both the pretraining and post-training phases, which their experiments show can significantly diminish the effects of these attacks while maintaining the models' effectiveness.

\subsubsection{Polymorphic and Metamorphic Malware}
\label{poly}

Polymorphic and metamorphic malware represent two sophisticated types of malicious software that are designed to evade detection. Polymorphic malware changes its code or signature patterns with each iteration, making it challenging for traditional signature-based detection methods to identify them consistently. In contrast, metamorphic malware goes a step further by not only changing its appearance but also altering its underlying code, essentially rewriting itself completely. This makes metamorphic malware even more elusive as it can vary its behavior and structure between infections. The advent of Generative AI and Large Language Models (LLMs) amplifies these threats significantly. These advanced AI systems can automate and refine the process of creating polymorphic and metamorphic malware, enabling them to generate numerous, highly variable, and sophisticated versions of malicious software at an unprecedented scale. This could potentially overwhelm conventional cybersecurity defenses, necessitating more advanced, AI-driven countermeasures to detect and neutralize these evolving cyber threats.

\subsection{Negative Externalities from the use of GenAI}\label{Externality}

Negative externalities are unintended, often harmful consequences of AI technologies that extend beyond the direct users or developers. In the context of generative AI, these externalities manifest as broader impacts on organizations and societies, such as data breaches at the firm level, where the sophistication of AI-driven attacks leads to increased costs, potential reputational damage, or the erosion of trust in digital systems, which weakens confidence in technology due to the pervasive use of AI-generated content and misinformation.

\subsubsection{Erosion of Trust in Digital Systems}

Trust is the cornerstone of our interactions with technology. Yet, as AI use expands, a troubling trend emerges—trust erosion in digital systems. This erosion is fueled by increased cyber threats, persuasive disinformation, privacy concerns tied to AI data collection, opaque algorithms, and regulatory gaps, among others. The European Union Agency for Law Enforcement Cooperation (Europol) highlights that "threat actors
will make increasing use of deepfake technology to facilitate various
criminal acts and conduct disinformation campaigns to influence or
distort public opinion." \footnote{ \href{https://www.europol.europa.eu/cms/sites/default/files/documents/Europol_Innovation_Lab_Facing_Reality_Law_Enforcement_And_The_Challenge_Of_Deepfakes.pdf}{EUROPOL - Law enforcement and the challenge of deepfakes}}

Beyond deepfakes, the rise of synthetic content which is forecast to increase as much as 90\% of the content online\footnote{\href{https://www.axios.com/2023/08/28/ai-content-flood-model-collapse}{Axios - AI could choke on its own exhaust as it fills the web}} is likely to further push trust levels towards media outlets down. The World Economic Forum\footnote{\href{https://www.weforum.org/agenda/2023/10/news-media-literacy-trust-ai/}{How can we build trustworthy media ecosystems in the age of AI and declining trust?
}} highlights that over 2 billion people will participate in elections in 2024 making the reliance on AI generated content central in the democratic processes. To ensure the thriving of our digital ecosystems, addressing this trust erosion, worsened by the proliferation of cyber shadows, is imperative.

\subsubsection{Data Breaches at the Firm Level}
\label{Data}

The development of GenAI represents a double-edged sword for the technology and business world. While AI can enhance the sophistication of cyber-attacks, target high-value data, and monitor cyber systems beyond traditional DevOps capabilities, it also increases the cost and sophistication of data breaches.
Since 2020, data breach costs have been rising steadily, with the global average in 2024 reaching \$4.88 million 
\footnote{\href{https://www.ibm.com/reports/data-breach}{IBM Security – Cost of a Data Breach Report}}. This increase is due to improvements in attack methods, greater exposure through extensive digital services, and the higher monetary value of compromised data. Before the introduction of data protection regulations, breach-related losses were asymmetric, as data-loss cyber incidents disproportionately impacted users compared to the financial impact on targeted firms. Rising consumer awareness has increased direct firm costs through reputational damage, trading partner losses, and supply-chain disruptions \cite{akey2021hacking}. However, the full shift of the financial burden wasn't fully realized until significant improvements in enforcement and potential fines were introduced with GDPR-like regulations \cite{koutroumpis2022under}. This situation is expected to put additional pressure on firms' finances and exacerbate the slowdown in firm dynamics, including entry and exit activities, due to higher compliance costs associated with GDPR-like regulations.

\subsubsection{Vulnerabilities in Critical Industries}

AI-driven threats are particularly disruptive to critical sectors such as financial services, healthcare, and energy infrastructure. These industries rely heavily on AI systems, making them vulnerable to large-scale cyberattacks that can compromise sensitive data and disrupt essential services. These industries serve as critical pillars of the digital economy, and disruptions in their operations have far-reaching effects on economic stability and societal trust.

\section{Toward Shadow Neutralization}

So, how do we neutralize those cyber shadows? In the quest to achieve cyber resilience, we find ourselves navigating through a complex landscape where no "silver bullet" solutions exist. The section delves into the intricate challenges and potential strategies for bolstering our digital defenses, with a primary focus on AI-driven security solutions and policy measures. While definitive answers may be elusive, the importance of exploring solutions and fostering widespread awareness cannot be overstated. This is crucial not only for safeguarding industries and infrastructures but also for protecting the global community and every individual within it. 

\subsection{AI-driven Threat (Shadow) Hunting}
\label{AIsolutions}

AI and ML technologies play a pivotal role in fortifying digital systems against a myriad of cyber threats \cite{schmitt2023securing}.
\footnote{https://www.weforum.org/agenda/2024/01/cybersecurity-ai-frontline-artificial-intelligence/} These technologies have been instrumental in developing sophisticated tools and methodologies for network intrusion detection, malware identification, spam filtering, and analyzing network traffic, among other applications. The integration of AI/ML in cybersecurity initiatives significantly enhances the ability to counteract emerging cyber threats through three main pillars: robustness, response, and resilience \cite{taddeo2019trusting}.

\textbf{Resilience:} AI's role extends to ensuring that cybersecurity systems can withstand attacks without significant compromise to their functionality. A crucial aspect of this resilience is the systems' adeptness at identifying potential threats through AI-driven cyber threat and anomaly detection mechanisms.

\textbf{Robustness:} AI-based cybersecurity systems are designed to maintain stability and continue functioning effectively even when targeted by adversarial attacks. They possess self-healing and self-testing capabilities, ensuring that the systems can recover from and adapt to various forms of cyber aggression.

\textbf{Response:} These systems are not just reactive but are also adaptive, learning from each incident to autonomously improve their defensive mechanisms. This includes the ability to launch countermeasures, create decoys, and establish honeypots to mislead and trap attackers, enhancing the overall security posture.

\subsubsection{Autonomous Threat Detection Systems}

Intrusion Detection Systems (IDS) safeguard computer networks and systems by detecting unauthorized access, breaches, and malicious activities, thus protecting data integrity, confidentiality, and availability \cite{schmitt2023securing}. IDS fall into two categories: 

\textbf{Network Intrusion Detection Systems (NIDS):} Scan network traffic for suspicious patterns indicating potential attacks.

\textbf{Host Intrusion Detection Systems (HIDS):} Are deployed on specific devices to monitor system and network operations, including changes in system logs and file modifications, pinpointing threats directly affecting the host.

ML is pivotal in cyber threat detection, employing various techniques to identify and mitigate a wide range of attacks. SVMs classify attacks like DoS and Probing, KNN reduces false alarms in intrusion detection, Decision Trees aid feature selection for network IDS, and, when combined with Genetic Algorithms, enhance IDS construction. Advanced methods, including Deep Learning with RNNs, LSTMs, and CNNs, excel at classifying anomalies and countering sophisticated attacks. AI in cyber threat detection reflects a growing reliance on data-driven approaches for robust cybersecurity. Intelligent Automated Cybersecurity solutions, harnessing these AI techniques, are essential for increasing the resilience and robustness of digital ecosystems against the threats posed by Generative AI.

AI will transform the landscape of cyber threat hunting by enabling more proactive and efficient identification of potential security threats or "cyber shadows". Through AI-driven threat hunting, organizations can leverage machine learning algorithms to sift through vast amounts of data at unprecedented speeds, identifying anomalies that could indicate a compromise or an attempted breach. This approach significantly reduces the time it takes to detect threats, moving from a reactive to a proactive stance in cybersecurity. AI algorithms can learn from past incidents, adapting to new tactics employed by cyber attackers, thereby continually improving threat detection capabilities. Moreover, AI-driven tools can automate the tedious and time-consuming tasks of data analysis, freeing up human analysts to focus on more complex investigations and decision-making processes. This integration of AI into cybersecurity operations enhances the ability to detect, analyze, and neutralize cyber shadows before they can manifest into full-blown cyber attacks, ensuring a more robust defense posture for organizations. However, full neutralization requires a response - an additional step beyond "threat detection", which is the execution of counter-measures / attacks either autonomous or via a human (see \ref{response}).

\subsubsection{Countering Malicious AI Image Alterations}

\cite{malicious_images_ICML23} describe a strategy to reduce the risks associated with malicious editing of images by large-scale diffusion models, which are advanced AI systems used for image manipulation. The central concept is to 'immunize' images, making them less vulnerable to being altered by these models. This is achieved by adding tiny, imperceptible changes to the images, known as adversarial perturbations, which are designed to interfere with how the diffusion models work. When these models try to manipulate an 'immunized' image, they end up creating unrealistic results. Additionally, the authors suggests a policy change for this approach to be truly effective. Instead of relying on individual users to implement this image protection, the responsibility should fall to the organizations that develop the diffusion models. These entities should actively support and carry out the immunization process. The strategy for mitigating risks from image editing by large-scale diffusion models using adversarial perturbations emphasizes cybersecurity robustness.

\subsubsection{Collaborative Intelligence in Cyber Threat Response}
\label{response}

Full neutralization of cyber threats requires a response that goes beyond mere threat detection. The response phase involves the execution of counter-measures or actions, which can be carried out autonomously by AI systems or in collaboration with human analysts \cite{human-in-the-loop_ICML23}. While automation is crucial for rapid response to known threats and routine tasks, the human element remains invaluable in complex investigations or more contextual decision making. It also ensures transparency, fairness, and therefore trust in AI-driven security solutions \cite{Trust_ICML23, fair-and-accurate_ICML23}.

\textbf{Automation} is critical for handling routine and repetitive tasks, especially in the early stages of threat detection and response. It can rapidly identify and block known threats, reducing the burden on human analysts. Automation also enables real-time monitoring and response, which is vital in the face of fast-evolving threats.

\textbf{Augmentation} enhances the capabilities of human analysts by providing them with AI-driven insights, recommendations, and data analysis tools. This collaboration allows analysts to make more informed decisions, investigate complex threats, and adapt to novel attack strategies effectively. Augmentation is particularly valuable when dealing with sophisticated attacks that may require human judgment and expertise.

Human AI Collaboration ensures that AI-driven responses are aligned with organizational goals and ethical considerations. Therefore, a balanced approach that combines automation with human expertise is the most effective strategy for risk assessment, detection, and response.

\subsection{Policy Measures to Alleviate Negative Externalities}
\label{Policy}

Striking the “right” balance between the cyber enhancements that GenAI can provide, and the new vulnerabilities introduced by its use, is not an easy task \cite{he2023large, taddeo2019trusting}. By making AI more accessible and democratized, we benefit from the cross-industry financial efficiencies but suffer from a wider threat of malicious actors appropriating the lower entry barriers. As the cost of acquiring and deploying AI capabilities decreases, a broader array of adversaries, even those with limited resources, can tap into sophisticated AI tools \cite{schmitt2023securing}. This means that the scale and frequency of AI-driven (phishing) attacks could surge, making it even more challenging for organizations to maintain an effective defense. The interplay of accessibility and affordability of AI technologies may intensify the threat landscape, underscoring the urgency for novel defensive measures.

This situation necessitates a delicate balance in policy making – regulatory and legislative efforts must offer sufficient protection and control without stifling AI's potential as an innovation accelerator. The European Union's AI Act marks a positive step towards this, yet feedback from the industry suggests that strong country-member opposition can curtail its reach. Meanwhile, the United States has implemented an AI Executive Order, which puts a strong emphasis on fostering innovation and does not fully address the complex cybersecurity concerns associated with AI.

To further mitigate the cybersecurity challenges posed by AI, several solutions are proposed. Firm-level regulation, akin to the GDPR but with specific focus on AI, is crucial (including the implementation of a risk-based framework) \cite{jain2023tuning}. Automated code improvement measures (like security hardening) should be integrated more robustly into the training of Large Language Models (LLMs) or “sit” on top of the AI assistant applications. As a result, techniques such as Reinforcement Learning from Compiler Feedback (RLCF) and controlled code generation are essential. These should be complemented by a standardized dataset of secure-coding practices to ensure AI systems are trained with cybersecurity in mind. Finally, regularly updating LLMs with the latest security vulnerabilities is vital to ensure they are equipped to identify and address emerging threats effectively \cite{he2023large}. This holistic approach, combining regulatory frameworks with advanced technological measures, is key to leveraging AI in cybersecurity while minimizing its potential risks.

\subsection{Cybersecurity Frameworks and AI Regulation}

Cybersecurity policies vary across regions, with the United States, Europe, and other global areas adopting different approaches to regulating AI and addressing digital threats.

In the United States, the recently signed AI Executive Order focuses on promoting innovation and ensuring that AI systems are safe, trustworthy, and uphold privacy and civil rights. The NIST Cybersecurity Framework, widely adopted across industries, provides a risk-based approach to managing cybersecurity threats, emphasizing detection, response, and recovery strategies in AI-based systems.

In Europe, the EU AI Act and General Data Protection Regulation (GDPR) work together to regulate AI systems, particularly those classified as high-risk. The EU AI Act mandates transparency, risk management, and human oversight in AI deployments, while GDPR ensures that personal data is processed with privacy by design and by default, protecting individual rights across the digital ecosystem.

In other regions, such as Asia, frameworks like Singapore's Model AI Governance Framework and Japan’s AI Strategy emphasize the responsible development of AI technologies. Singapore’s framework, for instance, promotes transparency and accountability in AI systems, while Japan's strategy focuses on fostering AI innovation while addressing ethical concerns and ensuring security.

\subsection{AI Risk Management in the EU Regulatory Framework}
\label{Scenario}

Under the EU AI Act, AI systems used for high-risk applications, such as fraud detection in financial services, must incorporate stringent safeguards to protect individual rights and ensure compliance with regulatory standards. In such cases, developers of AI-driven cybersecurity tools are required to implement comprehensive risk management processes. These systems must be designed with privacy by design and privacy by default principles, ensuring that personal data is processed securely and fairly.

For instance, an AI-based fraud detection system could monitor financial transactions in real time, identifying unusual patterns that indicate potential fraud. To comply with both the AI Act and GDPR, this system would include transparency measures, informing users that AI is involved in decision-making processes and allowing for human oversight to prevent errors or biases. Additionally, the system would need to ensure that users' privacy rights are upheld by minimizing the amount of personal data processed and securing it through robust encryption methods.

By integrating the requirements of the EU AI Act with GDPR’s privacy protections, such AI systems would mitigate risks while ensuring regulatory compliance and safeguarding individual freedoms within digital ecosystems.

\section{Discussion}

\subsection{Integrating Policy and Technology}

The most effective way to safeguard our digital ecosystems lies in a strategic blend of targeted policy measures and the deployment of AI-driven security technologies. This dual approach leverages the strengths of both proactive policy frameworks and advanced technological solutions to create a comprehensive defense mechanism against cyber threats.

\textbf{Targeted policy measures} are crucial for establishing the legal and regulatory framework necessary for cybersecurity. They set the standards and define expectations for behavior within digital spaces, enforce compliance, and deter malicious activities through legal repercussions. Policies can mandate essential security protocols, data protection requirements, and incident response strategies, ensuring that organizations have a baseline level of security that they must achieve.

\textbf{AI-driven security technologies}, on the other hand, provide the dynamic and adaptive capabilities needed to combat sophisticated cyber threats and practically enforce the agreed policy measures. AI and ML can analyze vast amounts of data at incredible speeds, identify patterns indicative of cyber attacks, and evolve in response to new threats. This technology enables real-time threat detection and response, going beyond the static defenses traditional security measures offer.

The combination of these two approaches allows for a robust cybersecurity posture that is both preventive, through policy, and reactive, through technology. For example, shifting the responsibility of implementing image protection from individual users to the organizations developing diffusion models adds a layer of systemic robustness. It ensures that the protective measures are consistently applied at the source, further strengthening the overall defense against malicious image editing \cite{malicious_images_ICML23}. 
Policies provide the framework and guidelines that shape the implementation and use of AI technologies, ensuring they are used responsibly and effectively.
However, policies can be "toothless" if left without the necessary enforcement capabilities. For the image protection example, new techniques including Glaze\footnote{https://glaze.cs.uchicago.edu/} - a "defense" method against style mimicry - and Nightshade\footnote{https://nightshade.cs.uchicago.edu/} - an "offense" tool to distort feature representations inside generative AI image models - can make this battle for image protection copyrights a lot smoother.

Still, it is important to recognize the challenges and complexities involved in harmonizing policy and technology. Policies may lag behind the rapid pace of technological innovation, and AI technologies may encounter ethical, privacy, or accuracy concerns that need careful regulation. Hence, continuous dialogue between policymakers, technologists, and cybersecurity professionals is essential to refine and adjust both policy measures and technology deployments, ensuring they remain effective in the face of evolving cyber threats.

\subsection{Looking Ahead}
\label{Future}
As we pivot from the intricacies of our current cybersecurity paradigm towards the future, it becomes clear that the nuanced approach combining targeted policy measures with AI-driven security technologies sets a foundational blueprint for advancing digital protection. Understanding the real trade-offs in this process is essential for the adoption of AI technologies in ways that increase social welfare. The journey ahead, while promising, is fraught with evolving challenges that necessitate not only adaptation but also anticipation. For example, the advent of AI and ML ushers in the potential for fully autonomous attack agents \cite{multi-agents_ICML23, Schmitt2024}, representing a formidable risk in the cybersecurity landscape. Unlike most previous technologies that primarily enhanced productivity, AI introduces the capacity for autonomy. This characteristic significantly amplifies the threat level, as AI can independently initiate, execute, and adapt cyber attacks without human intervention. Such autonomous agents could exploit vulnerabilities at speeds and complexities far beyond human capability to respond, making them a critical concern for future cybersecurity measures. This shift towards autonomous capabilities necessitates a proactive and sophisticated approach to cyber defense, emphasizing the development of equally advanced AI-driven security solutions that can anticipate, identify, and neutralize threats autonomously, ensuring a dynamic and resilient digital ecosystem.

\textbf{Evolution of AI Technologies:} The rapid pace of technological innovation mandates that AI-driven security solutions must not only respond to current threats but also adapt to anticipate future vulnerabilities (e.g., quantum computing and extended reality \cite{Qamar2023ASystems}). This involves leveraging advances in AI and machine learning to develop more sophisticated, predictive models that can preemptively identify and neutralize threats before they materialize.

\textbf{Dynamic Policy Frameworks:} As digital ecosystems evolve, so too must the policy frameworks that govern them. This requires a proactive approach to policy-making, where regulations are regularly reviewed and updated to reflect the latest technological advancements and threat vectors. Such dynamic policies must balance the need for security with the imperative of fostering innovation and protecting individual privacy.

The path to robust cybersecurity is iterative and collaborative, demanding ongoing innovation, adaptable policies, and stakeholder cooperation. This nuanced approach paves the way for secure, resilient, ethical, and inclusive digital ecosystems in the future.

\section{Conclusion
}
\label{Conclusion}

The remarkable capabilities of GenAI present a unique blend of potential benefits and risks. When used responsibly, AI can contribute significantly to addressing critical challenges, enhancing prosperity, productivity, innovation, and security. Conversely, GenAI can cast diverse cyber shadows over the global economy. While the regulatory and legislative initiatives represent some vital first steps, further collaborative efforts across the government, the private sector, academia, and the civil society are needed to steer AI away from the shadows and into the light to achieve predominantly beneficial outcomes. Currently, one of the most pressing concerns is the amplification of existing threats. AI, in its current state, has the potential to intensify the severity and frequency of cyber threats already present in our digital landscape. This escalation demands immediate attention and action from policymakers and corporate leaders. 

Looking ahead, the landscape of threats is poised to evolve further. We anticipate the emergence of new, sophisticated threats, such as fully automated cyberattack agents powered by AI. These advanced threats could operate with unprecedented efficiency and scale, likely posing a challenge for traditional defense mechanisms. We need a nuanced and multidimensional approach to cybersecurity, recognizing that neither policy nor technology alone are sufficient to protect digital ecosystems. Instead, it is their strategic integration that offers the best chance of securing our digital futures against increasingly sophisticated cyber threats. 

It is important to note that the true extent of these risks and their impact is still largely unknown, which underscores the need for ongoing research and analysis to fully understand and address these challenges. As such, it is imperative that politicians and boards of directors take AI seriously. Their role in shaping policies, strategies, and responses to these evolving cyber threats is crucial. Without their active engagement and foresight, the cyber shadows cast by GenAI could darken, impacting not just individual companies or sectors, but the global economy and society at large. The time to act is now, to harness the immense power of AI for the greater good and safeguard our collective future.


\bibliography{main}
\bibliographystyle{IEEEtran}

\newpage

\vspace{11pt}

\end{document}